\documentclass{article}
\usepackage{spconf,amsmath,graphicx}
\usepackage{url}
\usepackage{amsmath}
\usepackage{bbding}
\usepackage{multirow}
\usepackage{hhline}

\usepackage{graphicx}
\usepackage{multirow}
\usepackage{hyperref}
\title{IITKGP-ABSP Submission to LRE22: Language Recognition in Low-Resource Settings}
\name{Spandan Dey$^1$, Md Sahidullah$^2$, Goutam Saha$^1$}
\address{$^1$ABSP Lab, Department of E~\&~ECE, Indian Institute of Technology Kharagpur, India\\
$^2$Independent Researcher, West Bengal, India\\
\url{sd21@iitkgp.ac.in}, \url{sahidullahmd@gmail.com}, \url{gsaha@ece.iitkgp.ac.in}}
\begin{document}
%
\maketitle
\begin{abstract}
This is the detailed system description of the IITKGP-ABSP lab's submission to the NIST language recognition evaluation (LRE) 2022. The objective of this LRE (LRE22) is focused on recognizing 14 low-resourced African languages. Even though NIST has provided additional training and development data, we develop our systems with additional constraints of extreme low-resource. Our primary fixed-set submission ensures the usage of only the LRE 22 development data that contains the utterances of 14 target languages. We further restrict our system from using any pre-trained models for feature extraction or classifier fine-tuning. To address the issue of low-resource, our system relies on diverse audio augmentations followed by classifier fusions. Abiding by all the constraints, the proposed methods achieve an EER of $11.43\%$ and cost metric of $0.41$ in the LRE22 development set. For users with limited computational resources or limited storage/network capabilities, the proposed system will help achieve efficient LID performance.
\end{abstract}


\section{Introduction}
\label{sec:intro}
Spoken language identification (LID) systems are used as front-end systems in multilingual speech applications~\cite{ambikairajah2011language}. The LID system first predicts the language spoken in the input speech. Based on the prediction, the subsequent language-specific speech application is activated for efficient multilingual usage. As the trend of \emph{human-to-computer-interaction} (HCI) is rapidly shifting from textual typing to verbal, the relevance of developing efficient LID systems is increasing worldwide. In the era of \emph{deep neural network} (DNN), LID system development has greatly benefited due to large-scale standard corpora development. The \emph{NIST language recognition evaluation} (LRE) challenges~\cite{sadjadi20182017,greenberg20122011,martin20102009,martin2003nist} are one of the major sources of LID databases. The \emph{oriental language recognition} (OLR) challenge~\cite{li2020ap20}, open-source databases such as the \emph{Mozilla Common Voice}~\cite{mozilla}, \emph{VoxLingua107}~\cite{valk2021voxlingua107} are also providing standard databases for LID system development. 

According to the $23^{rd}$ edition of \emph{Ethnolouge}~\cite{ethno23}, there are over 7117 spoken languages in the world. However, considering the current state of LID corpora development, the large-scale LID data is confined to only a few numbers of languages. There are many other languages across the globe for which developing DNN-based state-of-the-art LID systems is challenging due to a lack of resources. Among them, there are several prominent languages in South Asia, South East Asia, and Africa, with more than a million speakers~\cite{dey2024towards,dey23_interspeech,duroselle21b_interspeech}. Therefore, developing modern speech-based applications and gadgets is of immense social and economic interest. The latest edition of the NIST LRE challenge (LRE22) focuses on recognizing 14 low-resourced languages spoken across different parts of the African continent. These languages are: \emph{Afrikaans} (afr-afr), \emph{Tunisian Arabic} (ara-aeb), \emph{Algerian Arabic} (ara-arq), \emph{Libyan Arabic} (ara-ayl), \emph{South African English} (eng-ens), \emph{Indian-accented South African English} (eng-iaf), \emph{North African French} (fra-ntf), \emph{Ndebele} (nbl-nbl), \emph{Oromo} (orm-orm), \emph{Tigrinya} (tir-tir), \emph{Tsonga} (tso-tso), \emph{Venda} (ven-ven), \emph{Xhosa} (xho-xho), \emph{Zulu} (zul-zul). The LRE22 is an important initiative for developing speech technologies for the different African languages. 

\begin{table*}[!t]
\centering
\caption{Data-split, number of utterances, and duration of data for the LRE22-dev database that is used in our submission.}
\label{table:database}
\resizebox{\linewidth}{!}{%
\begin{tabular}{|l|l|l|l|l|l|l|l|l|l|l|l|l|l|l|l|} 
\hline
\multicolumn{2}{|l|}{Language-code} & afr-afr & ara-aeb & ara-arq & ara-ayl & eng-ens & eng-iaf & fra-ntf & nbl-nbl & orm-orm & tir-tir & tso-tso & ven-ven & xho-xho & zul-zul \\ 
\hline
\multirow{2}{*}{\begin{tabular}[c]{@{}l@{}}Duration in \\hours-minutes (h-m)\end{tabular}} & Training & 1h-49m & 1h-40m & 1h-50m & 1h-49m & 1h-36m & 1h-29m & 1h-35m & 1h-34m & 2h-11m & 2h-00m & 1h-57m & 1h-36m & 1h-28m & 1h-56m \\ 
\cline{2-16}
 & Validation & 18m & 23m & 21m & 23m & 22m & 18m & 20m & 19m & 22m & 23m & 26m & 21m & 20m & 27m \\ 
\hline
\multirow{2}{*}{\begin{tabular}[c]{@{}l@{}}Total number\\of files\end{tabular}} & Training & 250 & 250 & 250 & 250 & 250 & 250 & 250 & 250 & 250 & 250 & 250 & 250 & 250 & 250 \\ 
\cline{2-16}
 & Validation & 50 & 50 & 50 & 50 & 50 & 50 & 50 & 50 & 50 & 50 & 50 & 50 & 50 & 50\\
\hline
\end{tabular}
}
\end{table*}

The evaluation plan of LRE22~\cite{LRE22,LRE22_plan} has two modes of participation: (i) in the mandatory fixed-set, participants are allowed to use only the specified databases for system development, (ii) in the optional but recommended open-set, participants can use any data of their choice for system development. The permitted databases for the fixed-set condition are the 2017 NIST LRE Development Set, the training data for the previous LRE editions, the 2017 NIST LRE Test Set, and the 2022 NIST LRE Development Set. Users can also use the VoxLingua107~\cite{valk2021voxlingua107} database. For augmentation, non-speech audio samples can also be used. Our primary submission, discussed in this work, is based on the fixed-set condition. Since we are focused on low-resourced languages, there can often be other accompanying limitations (especially by independent researchers or small organizations) regarding data availability, computational resources, or limited network capacity to download larger databases. Our system's key objective is to develop low-resource LID systems abiding by these additional constraints. Therefore, along with the existing rules of the LRE22 fixed-set condition, we strictly follow some other confinements in data usage and classifier training:
\begin{itemize}
    \item We only use utterances of the target languages (14 African languages). The 2022 NIST LRE Development Set is only used for system development. 
    \item Any pre-trained large architecture (even if trained using the permitted fixed-set databases) is not used as feature extractor or fine-tuning.
\end{itemize}

With the state-of-the-art large DNN-based architectures, using all the permitted databases can yield more efficient LID performance~\cite{snyder2018spoken}. Acknowledging that, our goal is to develop a LID system that can compete as close as possible with only a fraction of the resources. We develop several audio augmentation methods to create diverse domain information within the limited development data. After that, we develop multiple feature sets and classifiers to develop different LID systems. Finally, the LID systems' prediction scores are fused for final inference. Our work enlightens a unique and practical perspective on the issue of ``\emph{low-resource}" that coincides well with the theme of the current LRE challenge.

The rest of the paper is organized as follows: Section~\ref{sec:database} presents the detailed data organization, and Section~\ref{sec:system} is used for describing the proposed LID system. The experiment results and discussions are presented in Section~\ref{sec:results} followed by the conclusion.

\section{Summary of the database}
\label{sec:database}
\subsection{2022 NIST LRE Development Set: Data split}
Concerning extremely low-resourced conditions, our developed system aims to provide competitive LID results using only the 2022 NIST LRE Development Set (hereafter referred to as LRE22-dev). The LRE22-dec data contains only 4200 utterances of the 14 languages (300 each). Each utterance within a language is further collected from 30 different recordings (10 utterances from each). From the meta information and by manual inspection, it is likely that the recordings are collected independently from different speakers. We first create train and validation sets from the LRE22-dev data. Following the session information available in the meta-data, we randomly select the utterances from 25 recording sessions as training and the remaining five recording sessions as validation. Therefore, for each target language, our training data consists of only 250 utterances, and the validation data contains only the speaker-disjoint remaining 50 utterances. In Table~\ref{table:database}, we have summarized the details of the LRE22-dev data that is only utilized in our submission.

The final evaluation is conducted on the 2022 NIST LRE Evaluation Set (LRE22-eval). It contains 26,473 utterances belonging to the 14 target languages. Both the LRE22 dev and eval utterances are in $.\mathrm{sph}$ format and sampled in 8~kHz of the sampling rate. The duration of the utterances varies between 3s to 30s. To avoid data-split related bias, we also conduct our initial experiment (System-1, discussed in Section~\ref{sec:system}) with three different splits with 250 training and 50 validation session-disjoint utterances per language. The results show lower variation across splits. So, we randomly fix one of the random splits for the subsequent experiments and train the LID systems.

\section{System descriptions}
\label{sec:system}
The key approach of our submission is to explore diverse audio data augmentation methods and develop multiple end-to-end LID classifiers. The different LID classifiers are further developed using complementary speech features and state-of-the-art classifiers. Finally, we apply augmentation-fusion, feature-fusion, classifier-fusion, and overall fusion of all systems. Next, we have briefly described each of the systems we develop:
\subsection{LID system development details}
\label{sec:subsystem}
We first apply an energy-based voice activity detector (VAD) followed by framing the utterances using a window of 25ms and a hop length of 10ms. After the pre-processing, we extract 40-dimensional \emph{mel-frequency cepstral coefficient} (MFCC) features followed by utterance-level cepstral mean subtraction (CMS). The features are then segmented into chunks of 3s. The training utterance features are then used to train the classifiers. We primarily use TDNN-based classifiers along with one Gaussian mixture model (GMM) based generative classifier. X-vector~\cite{snyder2018x} classifier contains a TDNN encoder at the beginning with dilated convolution layers. After the encoder, there is a statistics pooling layer that converts variable lengths into fixed-dimension embedding. Finally, we use two fully connected layers followed by the softmax output layer. ResNet-based TDNN~\cite{kreyssig2018improved} uses the concepts of residual connections in the TDNN encoder of the x-vector. It is reported to improve the robustness and convergence of training. ECAPA-TDNN~\cite{desplanques2020ecapa} applies several modifications of the conventional x-vector time-delay neural network (TDNN)~\cite{snyder2018x} and is shown to outperform several advanced TDNN classifiers for different speech processing application. ECAPA-TDNN replaces the TDNN encoders of the x-vector with the squeeze-excitation residual block (SE-Res2), which increases the temporal context further. The SE-Res2 blocks are further connected with multi-layer feature aggregation (MFA), which increases the robustness of the classifier. Finally, the statistics pooling layer of the x-vector is replaced with channel-attentive pooling. 

The classifiers are developed in PyTorch. The training is done for 30 epochs, with angular-margin softmax loss function (hyper-parameters:$s=30,m=0.2$), AdamW optimizer, with an initial learning rate of $0.001$. The learning rate is scheduled by reduction by $0.5$ for one epoch if, for two consecutive epochs, the validation loss increases. We use the ``lre-scorer" script, which is provided by the LRE22 organizers, to compute the costs (actCprimary and minCprimary) of the system. Additionally, we use the equal error rate (EER) metric computed from the ASVsubtools~\footnote{\url{https://github.com/Snowdar/asv-subtools}}. The system development steps are kept the same for all the subsequent systems we describe next.

\subsection{Fusions based on augmentation}
To address the issue of low-resource, we rely on exploring diverse audio data augmentation approaches that can be segregated into three categories. In our previous works~\cite{dey2021cross,dey2021overview,dey2023cross}, we studied an extensive number of different audio augmentation methods across multiple corpora that also include part of the NIST LRE 2011 CTS data. Among them, we here select only those augmentation methods that showed consistent and satisfactory LID performances across several corpora. The augmentations we explore are: 
\begin{enumerate}
    \item \textbf{Additive non-speech perturbation}: Non-speech effects, such as additive noise, babble noise, and music clips are added to the LRE22-dev training utterances. We implement this augmentation with the Kaldi tool's~\cite{povey2011kaldi} VoxCeleb recipe~\footnote{\url{https://github.com/kaldi-asr/kaldi/tree/master/egs/voxceleb/v2}}. The non-speech samples are used from the MUSAN corpus~\cite{snyder2015musan}. Using the Kaldi's tool, we also apply convolution to the training utterances with randomly selected room impulse responses from the RIR dataset. The motivation for applying this augmentation is to improve system robustness in case the small training data is biased for specific kinds of background noises and room environments.
    \item \textbf{Perturbation of signal characteristics:} Here, rather than adding non-speech contents, we randomly tweak the signal volume, speed, and pitch. We have used the MATLAB audio degradation toolbox~\cite{matthias2013a} for implementing this augmentation category. The volume of the utterances is randomly tuned within $[-30dB,+40dB]$. For pitch perturbations, random shifts by $[-4,4]$ semitones are conducted without changing the utterance lengths. The speed is changed by randomly choosing a factor within $\Gamma \in [-15,+15]$. $\Gamma >0$ increases playback speed and $\Gamma <0$ reduces it. The motivation for this augmentation is to incorporate the different speaker characteristics in our small training set. It will help to improve robustness if the unseen test sample speakers possess different characteristics.
    \item \textbf{Applying random speech enhancements:} The MATLAB Voicebox tool~\cite{brookes1997voicebox} is used to randomly apply one of the three speech enhancement algorithms to the utterances: spectral subtraction~\cite{berouti1979enhancement}, $\log$ MMSE noise estimation based enhancement~\cite{gerkmann2011unbiased}, and de-reverberation~\cite{doire2016single}. This augmentation method is applied to ensure noise robustness.
\end{enumerate}

\begin{table*}
\centering
\caption{Details of all the systems we develop for the LRE22 submission.}
\label{tab:sys}
\resizebox{\linewidth}{!}{%
\begin{tabular}{|c|c|c|c|c|c|c|c|c|c|} 
\hline
\multirow{2}{*}{\textbf{System}} & \multicolumn{3}{c|}{\textbf{Augmentations }} & \multicolumn{2}{c|}{\textbf{Features }} & \multicolumn{4}{c|}{\textbf{Classifiers }} \\ 
\cline{2-10}
 & \textbf{Non-speech} & \textbf{Signal-perturbation} & \textbf{Enhancement} & \textbf{MFCC} & \textbf{RASTA-PLP} & \textbf{X-vector TDNN} & \textbf{ECAPA-TDNN} & \textbf{ResNet-TDNN} & \textbf{Statistical Modeling} \\ 
\hline
S0 & \XSolidBrush & \XSolidBrush & \XSolidBrush & \Checkmark & \XSolidBrush & \Checkmark & \XSolidBrush & \XSolidBrush & \XSolidBrush \\ 
\hline
S1 & \XSolidBrush & \XSolidBrush & \XSolidBrush & \Checkmark & \XSolidBrush &  \XSolidBrush & \Checkmark & \XSolidBrush & \XSolidBrush \\ 
\hline
S2 & \Checkmark & \XSolidBrush & \XSolidBrush & \Checkmark & \XSolidBrush &\XSolidBrush & \Checkmark & \XSolidBrush & \XSolidBrush \\  
\hline
S3 & \XSolidBrush & \Checkmark & \XSolidBrush & \Checkmark & \XSolidBrush & \XSolidBrush & \Checkmark & \XSolidBrush & \XSolidBrush \\  
\hline
S4 & \XSolidBrush & \XSolidBrush & \Checkmark & \Checkmark & \XSolidBrush & \XSolidBrush & \Checkmark & \XSolidBrush & \XSolidBrush \\ 
\hline
S5 & \Checkmark & \Checkmark & \Checkmark & \Checkmark & \XSolidBrush & \XSolidBrush & \Checkmark & \XSolidBrush & \XSolidBrush \\ 
\hline
S6 & \Checkmark & \Checkmark & \Checkmark & \XSolidBrush & \Checkmark & \XSolidBrush & \Checkmark & \XSolidBrush & \XSolidBrush \\ 
\hline
S7 & \Checkmark & \Checkmark & \Checkmark & \Checkmark & \XSolidBrush & \XSolidBrush & \XSolidBrush & \Checkmark & \XSolidBrush \\
\hline
S8 & \XSolidBrush & \XSolidBrush & \XSolidBrush & \Checkmark & \XSolidBrush & \XSolidBrush & \XSolidBrush & \XSolidBrush & \Checkmark \\
\hline

\end{tabular}
}
\end{table*}

\begin{table*}[!t]
\centering
\caption{LID system performance on the LRE22-dev validation utterances for the different fusion-systems in our submission. Here. ``+" denotes the score fusion among the systems.}
\label{tab:final}
\resizebox{.7\linewidth}{!}{%
\begin{tabular}{|c|c|c|c|c|} 
\hline
\textbf{Fusion mode} & \textbf{System} & \textbf{actCprimary} & \textbf{minCprimary} & \textbf{EER (\%)} \\ 
\hhline{|=====|}
\multirow{2}{*}{No-augmentation} & S0 & 0.73588 & 0.62787 & 21.42 \\ 
\cline{2-5}
 & S1 & 0.66648 & 0.57703 & 20.82 \\ 
 \hhline{|=====|}
\multirow{7}{*}{Augmentation} & S2 & 0.68632 & 0.60709 & 19.71 \\ 
\cline{2-5}
 & S3 & 0.69011 & 0.58687 & 17.71 \\ 
\cline{2-5}
 & S4 & 0.62692 & 0.54236 & 21.14 \\ 
\cline{2-5}
 & S1+S2 & 0.54593 & 0.53495 & 16.00 \\ 
\cline{2-5}
 & S1+S3 & 0.54593 & 0.53495 & 16.00 \\ 
\cline{2-5}
 & S1+S4 & 0.56952 & 0.55455 & 16.35 \\ 
\cline{2-5}
 & S1+S2+S3+S4 & 0.54236 & 0.53571 & 15.84 \\ 
\hhline{|=====|}
\multirow{4}{*}{Feature} & S5 & 0.54890 & 0.49819 & 19.45 \\ 
\cline{2-5}
 & S6 & 0.70319 & 0.62830 & 16.42 \\ 
\cline{2-5}
 & S5+S6 & 0.44258 & 0.43599 & 16.55 \\ 
\hhline{|=====|}
\multirow{3}{*}{Classifier} & S5 & 0.54890 & 0.49819 & 19.45 \\ 
\cline{2-5}
 & S7 & 0.70319 & 0.62830 & 21.46 \\ 
\cline{2-5}
 & S5+S7 &  0.52511b& 0.51802 & 15.14 \\
\hhline{|=====|}
 Statistical model & S8 & 0.74154 & 0.68148 & 25.14 \\
\hhline{|=====|}
  Final fusion & S0+S1+S2+S3+S4+S5+S6+S7+S8 & \textbf{0.41632} & \textbf{0.41385} & \textbf{11.43} \\
\hline
\end{tabular}
}
\end{table*}

\begin{table*}[!t]
\centering
\caption{Language-wise EER (\%) for all the developed LID systems.}
\label{table:lang-wise}
\resizebox{\linewidth}{!}{%
\begin{tabular}{|l|l|l|l|l|l|l|l|l|l|l|l|l|l|l|} 
\cline{2-15}
\multicolumn{1}{l|}{} & \textbf{afr-afr} & \textbf{ara-aeb} & \textbf{ara-arq} & \textbf{ara-ayl} & \textbf{eng-ens} & \textbf{eng-iaf} & \textbf{fra-ntf} & \textbf{nbl-nbl} & \textbf{orm-orm} & \textbf{tir-tir} & \textbf{tso-tso} & \textbf{ven-ven} & \textbf{xho-xho} & \textbf{zul-zul} \\ 
\hline
S0 & 12.00 & 16.07 & 24.00 & 46.00 & 40.00 & 18.00 & 22.00 & 20.00 & \textbf{0.00} & 0.30 & 27.84 & 18.00 & 20.00 & 33.84 \\ 
\hline
S1 & 10.00 & 24.00 & 22.66 & 27.51 & 24.60 & 26.00 & 23.71 & 23.25 & 21.10 & 19.80 & 20.36 & 19.83 & 19.84 & 20.28 \\ 
\hline
S2 & 7.77 & 22.05 & 20.31 & 32.08 & 22.00 & \textbf{12.08} & \textbf{5.62} & \textbf{16.00} & \textbf{0.00} & 2.00 & 26.00 & 18.15 & 21.85 & \textbf{16.56} \\ 
\hline
S3 & \textbf{4.07} & 15.00 & \textbf{13.33} & \textbf{18.11} & \textbf{20.40} & 19.28 & 18.01 & 18.00 & 17.10 & 16.20 & \textbf{16.18} & \textbf{17.00} & 16.92 & 17.17 \\ 
\hline
S4 & 6.00 & 18.00 & 19.84 & 27.50 & 27.60 & 26.33 & 25.42 & 24.30 & 22.44 & 20.62 & 20.88 & 21.33 & 21.38 & 21.14 \\ 
\hline
S5 & 6.00 & 15.00 & 14.00 & 23.53 & 25.20 & 24.30 & 22.85 & 22.25 & 20.22 & 19.20 & 18.95 & 19.16 & 19.26 & 19.45  \\ 
\hline
S6 & 8.38 & \textbf{7.88} & 14.66 & 22.50 & 22.80 & 20.66 & 19.71 & 19.27 & 17.53 & 16.83 & 16.54 & 16.29 & \textbf{16.58} & 16.42 \\ 
\hline
S7 & 12.46 & 17.00 & 20.00 & 20.96 & 23.20 & 23.33 & 23.36 & 22.50 & 20.61 & 19.16 & 20.90 & 20.83 & 21.23 & 21.46 \\ 
\hline
S8 & 14.00 & 21.53 & 28.00 & 42.00 & 47.69 & 38.00 & 22.00 & 20.07 & \textbf{0.00} & \textbf{0.07} & 46.38 & 18.00 & 20.00 & 20.00  \\
\hline
\end{tabular}
}
\vspace{-.3cm}
\end{table*}
\begin{figure*}[!t]
\vspace{-.1cm}
    \centering
    \includegraphics[trim=0cm 0cm 0cm 0cm,clip,width=.72\textwidth]{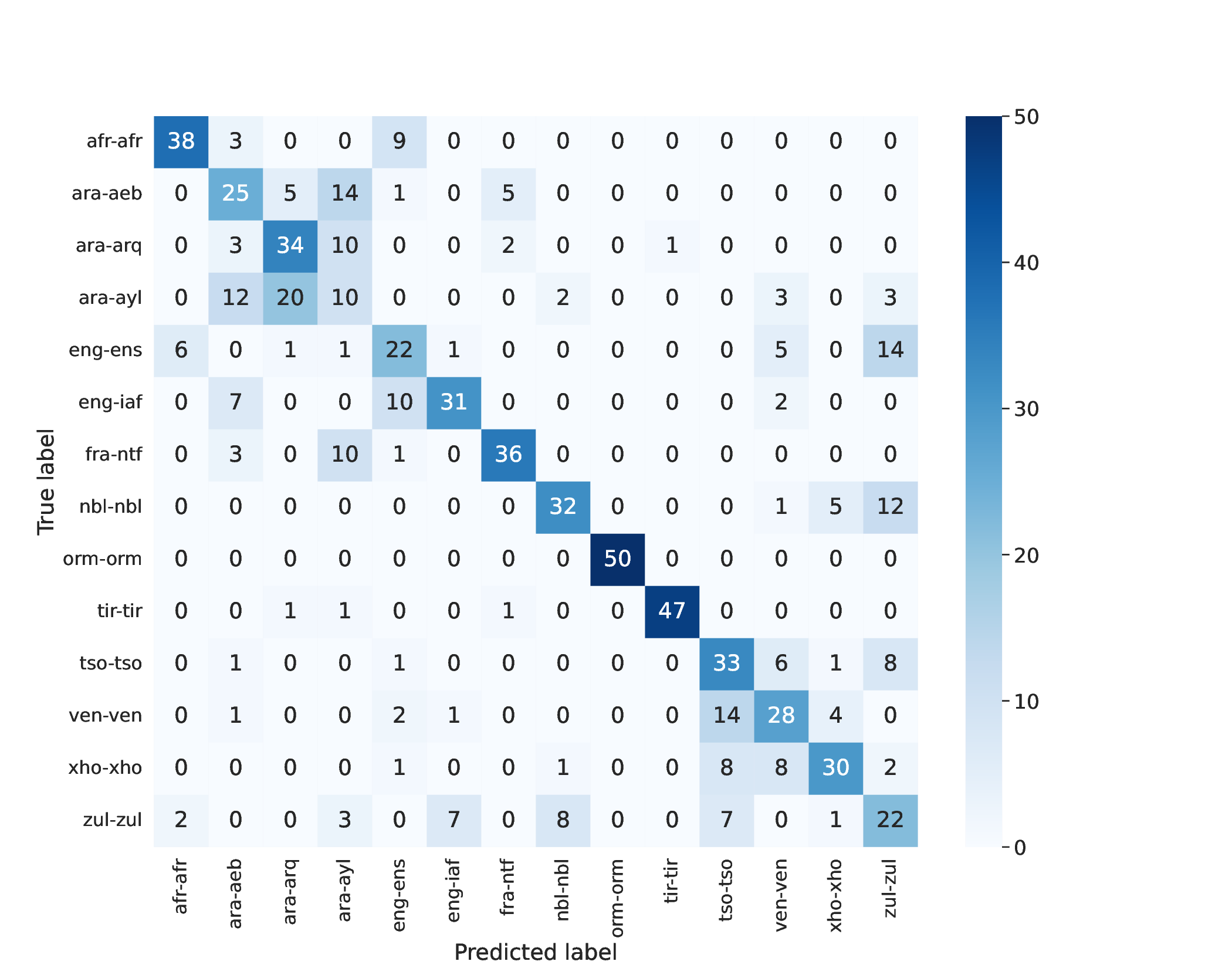}
    \vspace{-.25cm}
    \caption{Confusion matrix of final fused system's prediction for the LRE22-dev validation set.}
    \label{fig:confusion}
    \vspace{-.1cm}
\end{figure*}
For each augmentation category, we train independent ECAPA-TDNN LID classifiers (System-2, System-3, and System-4). The prediction scores on the LRE22-dev validation set utterances are then used for computing the weights of the logistic regression-based score fusion using the MATLAB multi-focal tool.~\footnote{\url{https://sites.google.com/site/nikobrummer/focalmulticlass}}. For the subsequent fusion modes (feature and classifier), we follow the same fusion strategy.

\subsection{Fusions based on features}
From the three kinds of augmentation, we sample the training utterances and create pooled-augmented data such that it contains 7000 utterances (twice the number of training utterances).
With the ECAPA-TDNN classifier and the pooled augmented training set (total of 10500 utterances), we train two LID systems and use their predictions for score fusion. The first one (system-5) is the default 40-dimensional MFCCs. We use MFCCs as default because they are one of the most commonly used speech features across different speech processing tasks. However, in the LID literature, it is reported by researchers that \emph{perceptual linear predictive} (PLP) features followed by RASTA~\cite{hermansky1994rasta} post-processing worked better than MFCCs~\cite{ambikairajah2011language,dey2021overview}. So, for the second one (System-6), we extract 20-dimensional RASTA-PLP features.~\footnote{\url{https://www.ee.columbia.edu/~dpwe/resources/matlab/rastamat/}}.

\subsection{Fusions with classifiers}

As discussed in Section~\ref{sec:subsystem}, we train separate LID systems with three TDNN architectures (discriminative) and a GMM classifier (generative). The three TDNN classifiers are: (1) x-vector TDNN~\cite{snyder2018spoken}, (2) ECAPA-TDNN~\cite{desplanques2020ecapa}, and (3) ResNet-TDNN~\cite{kreyssig2018improved}. The x-vector TDNN is the basic system (System-0) that we use. Here, we train this model without any augmentation. In System-1, the ECAPA-TDNN model is used to train over the non-augmented LRE22-dev validation data. Later, we create the pooled-augmented data (discussed above) to train ECAPA-TDNN (System-5) and ResNet (System-7) systems. 

Apart from the discriminative classifier, we also train a system with a GMM-based generative classifier (System-8) on the non-augmented original training data and thereafter apply it in score fusion. This system uses 16-dimensional MFCCs augmented with 112-dimensional shifted delta coefficients (SDCs) as used in~\cite{duroselle21b_interspeech}. We train the language-specific GMM models by adapting the mean of the universal background model (UBM) trained with 2048 mixtures from our entire training set. 

In Table~\ref{tab:sys}, we have summarized all the different LID systems that we have developed for the final fusion.

\section{Results}
\label{sec:results}
\subsection{Fusion experiments}
For each augmentation-fusion experiment, we develop three LID systems. Thereafter, for the feature-fusion experiments, we develop two LID systems. Finally, for the classifier-fusion experiments, we develop three LID systems. We have conducted fusion within each mode first to evaluate the effectiveness of score-fusions with different parameters. Finally, we combine the prediction scores of all nine systems to produce the final scores. The final fusion weights are computed on the LRE22-dev validation data and are then applied for the 2022 NIST LRE Evaluation Set. The systems we develop are complementary (can be seen from Table~\ref{table:lang-wise}), and hence fusing them leads to an EER improvement of $9.99\%$ from the baseline.
\subsection{Language-specific analysis}
For our final system, we also compute the language-specific performance analysis and present it in Table~\label{tab:lang}. This additional experiment is done to assess the relative difficulty level in recognizing different languages. Following Table~\ref{tab:lang}, we can see that some languages, such as ``ara-ayl", ``eng-iaf", ``tso-tso" are showing inferior performances across multiple systems. Whereas, ``afr-afr" is performing well across the systems. All other languages are showing a diverse range of performances for the different systems. This observation reveals the complementary nature of the different systems, motivating us for the fusion stages. To get further insight, in Figure~\ref{fig:confusion}, we present the confusion matrix of LRE22-dev validation set for the final system. The figure shows that dialects of Arabic across African nations are frequently getting confused. The same observation holds for the English styles provided in the development data. Other African-originated languages, such as ``afr-afr", "orm-orm", and ``tir-tir" are very rarely getting confused with others. So, mutual similarity and influence among the languages play an important role in the development set.

\subsection{Other implementation details}
Following the rules of system-description submission~\cite{LRE22_plan}, we also provide the time taken to compute the final evaluation scores along with the CPU memory. These results are presented in Table~\ref{tab:details}. The computation speed shows the ease of deploying the models in real-time applications. Of course, fusing multiple systems increases the prediction time at the cost of reliable inference. But here, we can see that even after using nine systems, generating a prediction of each utterance takes approximately less than half a second. Our focus is to deal with the situation of the extremely low availability of speech utterances. The results also show that our system can be developed even with very limited computation resources. With all these constraints, it is ultimately the performance that matters. Our system achieves an EER of $11.43\%$, which indicates satisfactory LID performance. In the future, our goal is to performance further.

\begin{table}
\centering
\caption{System and performance specification for computing prediction scores over the LRE22 evaluation set.}
\label{tab:details}
\resizebox{\linewidth}{!}{%
\begin{tabular}{|c|c|c|} 
\hline
\begin{tabular}[c]{@{}c@{}}\textbf{System }\\\textbf{specifications}\end{tabular} & \begin{tabular}[c]{@{}c@{}}\textbf{Evaluation-set prediction}\\\textbf{scores per 100 files}\end{tabular} & \begin{tabular}[c]{@{}c@{}}\textbf{CPU memory}\\\textbf{used }\end{tabular} \\ 
\hline
\multicolumn{1}{|c|}{Quadro RTX 5000} & \multicolumn{1}{c|}{40.15 seconds} & \multicolumn{1}{c|}{2.1 GB} \\
\hline
\end{tabular}
}
\end{table}

\section{Conclusions}
\label{sec:conclusions}

Our submission in the NIST LRE22 is focused on developing competitive LID systems under an extremely limited amount of data. Along with the LRE22 fixed-set conditions, we further put a constraint of using only the development data for training. The development data contains only 300 utterances per 14 target languages. To deal with the issue of low-resource, we rely on data augmentation and fusion. It will be interesting to check the overall standing of our submission with the state-of-the-art that can exploit a much larger amount of labeled or unlabelled audio data. The developed system is important as its orientation is different. If we can make LID systems with competitive performance with very limited data, then the ease of implementation can be largely improved. More numbers of languages across the globe can be utilized even without the initiation of large-scale corpus development.

\bibliographystyle{IEEEbib}
\bibliography{main}

\end{document}